\documentclass[%
 reprint,
 amsmath,amssymb,
 aps,
 prl,
]{revtex4-2}

\usepackage{graphicx}% Include figure files
\usepackage{dcolumn}% Align table columns on decimal point
\usepackage{bm}% bold math
\usepackage{color}
\usepackage{cleveref}
\usepackage{afterpage}
\usepackage{float}
\usepackage{amsmath}
\usepackage{bm}
\usepackage{lipsum}

\begin{document}
%\begin{widetext}
\preprint{APS/123-QED}

\title{Microswimmers knead nematics into cholesterics}

\author{Bhavesh Gautam}
\affiliation{%
 Univ. Bordeaux, CNRS, LOMA, UMR 5798, F-33400 Talence, France 
}

\author{Juho S. Lintuvuori}%
\affiliation{%
 Univ. Bordeaux, CNRS, LOMA, UMR 5798, F-33400 Talence, France 
}%

\date{\today}% It is always \today, today,
  
             \begin{abstract}
%\lipsum[1-1]
The hydrodynamic stresses created by active particles can destabilise orientational order present in the system. This is manifested, for example, by the appearance of a bend instability in active nematics or in quasi-2-dimensional living liquid crystals consisting of swimming bacteria in thin nematic films. Using large-scale scale hydrodynamics simulations, we study a system consisting of spherical microswimmers within a 3-dimensional nematic liquid crystal. We observe a spontaneous chiral symmetry breaking, where the uniform nematic state is kneaded into a continously twisting state, corresponding to a helical director configuration akin to a cholesteric liquid crystal. The transition arises from the hydrodynamic coupling between the liquid crystalline elasticity and the swimmer flow fields, leading to a twist-bend instability of the nematic order. It is observed for both pusher (extensile) and puller (contractile) swimmers. Further, we show that the liquid crystal director and particle trajectories are connected: in the cholesteric  state the particle trajectories become helicoidal.
\end{abstract}
\maketitle


\paragraph*{Introduction.}
Active materials consists of systems where the individual building blocks convert energy into work locally~\cite{ramaswamy2010mechanics}. Examples of this are provided by bacterial fluids~\cite{aranson2022bacterial}, catalytic Janus colloids~\cite{ebbens2018catalytic} or active microtubules~\cite{sanchez2012spontaneous}, at the micrometer length-scale. One of the striking features of these materials, is the emergence of collective motion on the scale considerably larger than the particles themselves, such as spontaneous formation of polar flocks in active colloids~\cite{bricard2013emergence} or the emergence of bacterial turbulence~\cite{dombrowski2004self}. 
An interesting subset of active materials is provided by active nematic gels~\cite{doostmohammadi2018active}. These consists of active units, force dipoles, with overlaying orientational, nematic, order. Pioneering work showed, using linear stability analysis, that the (active) force-dipoles can destabilise {\color{black} their} nematic order via hydrodynamic instabilities~\cite{simha2002hydrodynamic,ramaswamy2007active,voituriez2005spontaneous}. 

Another example is provided by finite size microswimmers moving in orientationally ordered fluids, {\color{black} where the flow-fields created by the swimmers interact with the topology of the surrounding fluid. Typical experimental} examples include  rod-like bacteria swimming in nematic liquid crystals crystals~\cite{zhou2014living,mushenheim2014dynamic,smalyukh2008elasticity,mushenheim2014using,zhou2017dynamic,chi2020surface,goral2022frustrated}. In the experiments, the bacteria is observed to align along the nematic director~\cite{smalyukh2008elasticity,goral2022frustrated} and the directed motion can be used to, for example, to transport cargo~\cite{mushenheim2014dynamic}. Recent experiments have shown that the LC topology can be used to control the swimmers~\cite{peng2016command,genkin2017topological,turiv2020polar,koizumi2020control}, such as trapping the particles with topological defects~\cite{peng2016command} or using LC patterns to create bacterial jets~\cite{turiv2020polar}, where collective (hydrodynamic) effects play a key role.

The swimming bacteria stir the surrounding fluid which can reorient the nearby nematic. In the simplest case of a uniform nematic LC, experiments in thin, quasi 2-dimensional, films  have demonstrated an orientational instability of the nematic order when bacterial activity is increased ~\cite{zhou2014living}. The coupling between the (collective) hydrodynamic effects created by the swimmers and the liquid crystalline elasticity, leads to a bend-instability of the LC director~\cite{zhou2014living}, similarly to what is predicted for extensile active nematic gels~\cite{simha2002hydrodynamic,doostmohammadi2018active} in 2-dimensions.

In this work, we open the 3rd dimension and consider microswimmer inclusions in a fully 3-dimensional nematic liquid crystal. By using hydrodynamic simulations we study the (collective) dynamics of spherical squirmers in the 3D sample. Our simulations reveal, an instability of the uniform nematic order, and a spontaneous formation of a continuous twist is observed. At the steady state, the LC director shows a constant twist along a unique axis, akin to a cholesteric state in passive LCs and the swimmer trajectories become helicoidal. This spontaneous chiral symmetry breaking arises from the coupling between the swimmer flow-fields and the nematic director. There is no prescribed chirality in the system, and indeed, on average, we observe the formation of right and left handed helices at approximately equal probabilities.  By evaluating the elastic distortions, we show that the spontaneous formation of the continuous twist can be understood in terms of a hydrodynamic twist-bend instability in 3-dimensions. 

\paragraph*{Model.} 
We use a lattice Boltzmann (LB) method to simulate the dynamics of microswimmers in liquid crystals~\cite{lintuvuori2017hydrodynamics,ludwigcode}.
The nematic LC is modelled using a Landau -- de Gennes free-energy whose density can be expressed 
\begin{align}\label{eq:fed}
F(Q_{\alpha\beta}) &= A_0\left(1-\frac{\gamma}{3}\right)\frac{Q_{\alpha \beta}^2}{2}-\frac{\gamma}{3}Q_{\alpha \beta}Q_{\beta \gamma}Q_{\gamma \alpha} \nonumber \\ 
& + \frac{\gamma}{4}(Q_{\alpha \beta}^2)^2 + \frac{K}{2}(\partial_{\beta}Q_{\alpha \beta})^2.
\end{align}
The Greek indices denote Cartesian coordinates and summation over repeated indices is implied.  $\mathbf{Q}$ is symmetric and traceless order parameter tensor, $A_0$ is a free energy scale, $\gamma$ is a temperature-like control parameter giving a order/disorder transition at $\gamma\sim 2.7$, and $K$ is an elastic constant. 

The evolution of $\mathbf{Q}$ is given by  the hydrodynamic Beris-Edwards equation~\cite{beris1994thermodynamics}
\begin{equation}
(\partial_t + u_{\nu}\partial_{\nu})Q_{\alpha \beta} - S_{\alpha\beta}= \Gamma H_{\alpha \beta}.
\end{equation}
where the first part describes the advection by velocity $\mathbf{u}$ and $S_{\alpha\beta}$ describes the possible rotation/stretching of $\mathbf{Q}$ by the flow~\cite{beris1994thermodynamics}. $\Gamma$ is the rotational diffusion constant and
the molecular field is given by
\begin{equation}
H_{\alpha \beta}= -{\delta 
  {\cal F} / \delta Q_{\alpha \beta}} + (\delta_{\alpha \beta}/3) {\mbox {\rm Tr}}({\delta {\cal F} / \delta Q_{\alpha \beta}}).
\end{equation}

To simulate the dynamics of the swimmers, we use a squirmer model~\cite{lighthill1952squirming}. The tangential (slip) velocity profile at the particle surface is given by~\cite{magar2003nutrient}
\begin{equation}
  u(\theta)= B_1\sin (\theta) + \frac{1}{2} B_2\sin(2\theta)
  \label{eq:veltangential}
\end{equation}
where $B_1$ and $B_2$ are constant, giving the strength of the source and force dipoles, respectively, and $\theta$ is the polar angle with respect to the particle axis ~\cite{ishimoto2013squirmer}. The source dipole sets the particle swimming speed $u_0=\frac{2}{3}B_1$ and the ratio $\beta=\frac{B_2}{B_1}$ is the squirmer parameter. 
In the LB method a no-slip boundary condition can be achieved by employing a bounce-back on links method (BBL)~\cite{ladd1994numerical1,ladd1994numerical2}, which needs to be modified for a moving surface~\cite{nguyen2002lubrication}. These local rules can include additional terms, such as a surface slip velocity (Eq.~\ref{eq:veltangential}) leading to LB simulations of squirming motion~\cite{llopis2010hydrodynamic,pagonabarraga2013structure}.

The fluid velocity obeys continuity equation, and the Navier-Stokes equation, which is coupled to the LC via a stress tensor~\cite{cates2009lattice}. We employ a 3D lattice Boltzmann algorithm to solve the equations of motion  using the Ludwig code~\cite{ludwigcode}.

{\it Simulation parameters:} We consider both pushers ($\beta < 0$) and pullers ($\beta > 0$). We fix the  $B_1=0.0015$, giving the particle velocity $u_0\equiv \tfrac{2}{3}B_1=10^{-3}$ in lattice units (LU), but vary the force dipole strength $B_2$ such that $\beta \in [-5,+5]$. We fix the  fluid viscosity $\eta = 0.167$ and the swimmer radius $R=6$ in LU. To model the nematic liquid crystal we use: $A_0 = 0.1$, $\gamma = 3.0$, $K = 0.005$, $\xi = 0.7$, $\Gamma = 0.3$ and a rotational viscosity $\gamma_1=\tfrac{2(3 s/2)^2}{\Gamma} = 5/3$, where $s$ is the scalar order parameter of the nematic. The physics of our system is governed by the Reynolds (Re) and Ericksen (Er) numbers, which give the ratio of inertial and viscous forces, as well as the ratio of viscous and elastic forces, respectively. Using the parameters above, we recover $\mathrm{Re} \equiv \tfrac{u_0R}{\eta}\approx 0.036$ and $\mathrm{Er}\equiv \tfrac{\gamma_1u_0R}{K}\approx 2$.
All the simulations were carried in a rectangular simulation box {\color{black}$21R\times 21R\times 21$, with periodic boundary conditions (PBCs) throughout.}

\paragraph*{Results:}

\begin{figure}
\includegraphics[width=1.0\columnwidth]{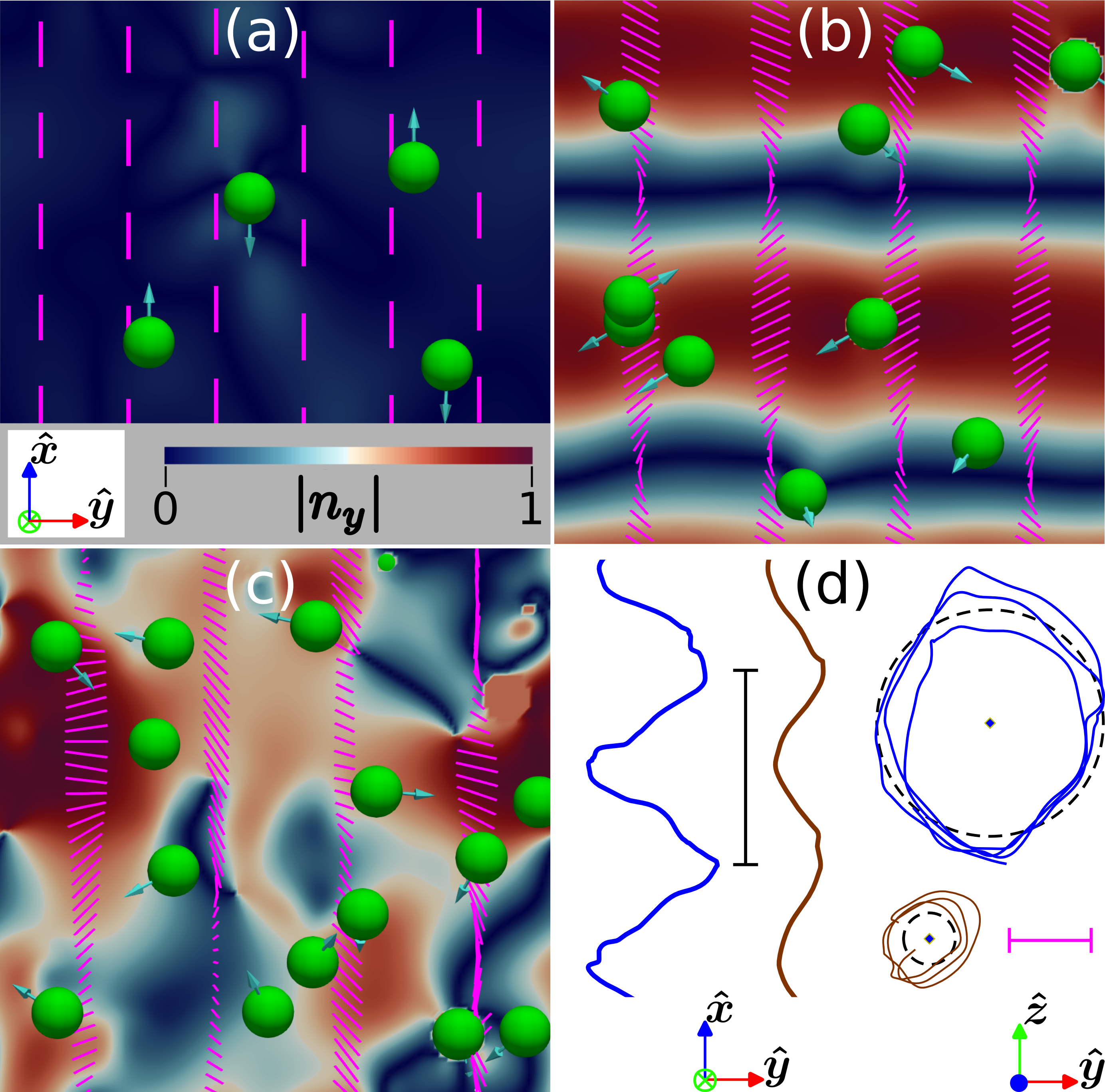}

\caption{(a-c) Examples of observed states in microswimmer nematic LC composites. (a)  At low volume fraction the system is uniform nematic and pushers (pullers) swim along (perpendicular) to the nematic director $\hat{\mathbf{n}}$. (b) When the activity of the system is increased, the uniform nematic becomes unstable, and a continuously twisting state is observed. The $\hat{\mathbf{n}}$ has a continuous twist along an unique axis ($x$-axis in this case). (c) At high activities, the spatial variations of $\hat{\mathbf{n}}$ become 3-dimensional leading to the formation of topological defects. (d) Examples of the unwrapped particle trajectories in the helical state, in the plane along (left) and perpendicular (right) to the helical axis, for pushers (blue lines) and pullers (brown lines), corresponding to $\phi\approx 0.01$ and $\beta=\pm 3.5$, respectively. The dashed lines corresponds to a theoretical argument (see text for details). The brown scale bar on the left corresponds to the system size $L\approx 21R$ and the pink on the right to $6R$. The data in (a)-(c) corresponds to $\beta\approx -2.0, -2.0, -4.5$ and $\phi \approx 0.01, 0.02, 0.02$, the background is colour coded according to $|n_y|$ and the nematic director is schematically shown by purple lines.}
\label{fig1}
\end{figure}

To study the collective dynamics of microswimmers in a 3-dimensional nematic liquid crystal, we initialised the system in a uniform nematic state with the $\hat{\mathbf{n}}$ along the $x$-axis (Fig.~\ref{fig1}a). The microswimmers were randomly distributed and oriented, while their volume fraction $\phi=\tfrac{N 4/3\pi R^3}{L_xL_yL_Z}$, and the strength of the force dipole $B_2$ and thus $\beta$ were varied. For a low $\phi$ and a low $|\beta|$ the system remain in a uniform nematic state, and pushers (pullers) have linear trajectories parallel (perpendicular) to the nematic director $\hat{\mathbf{n}}$ (Fig.~\ref{fig1}a) in agreement with the simulations of isolated swimmers~\cite{lintuvuori2017hydrodynamics}. When the global activity is increased, either by increasing $\phi$ or the magnitude of $\beta$, the uniform nematic becomes unstable, and the spontaneous formation of a cholesteric twist is observed (Fig.~\ref{fig1}b). At the steady state the $\hat{\mathbf{n}}$ twist continuously around a unique axis, and the particle trajectories become helicoidal (Fig.~\ref{fig1}d). Finally, the system looses the cholesteric order at higher activities. The director field variations lack a clear spatial symmetry and the particle dynamics become chaotic (Fig.~\ref{fig1}c)~\cite{SupMat}.

\begin{figure}
\includegraphics[width=1.0\columnwidth]{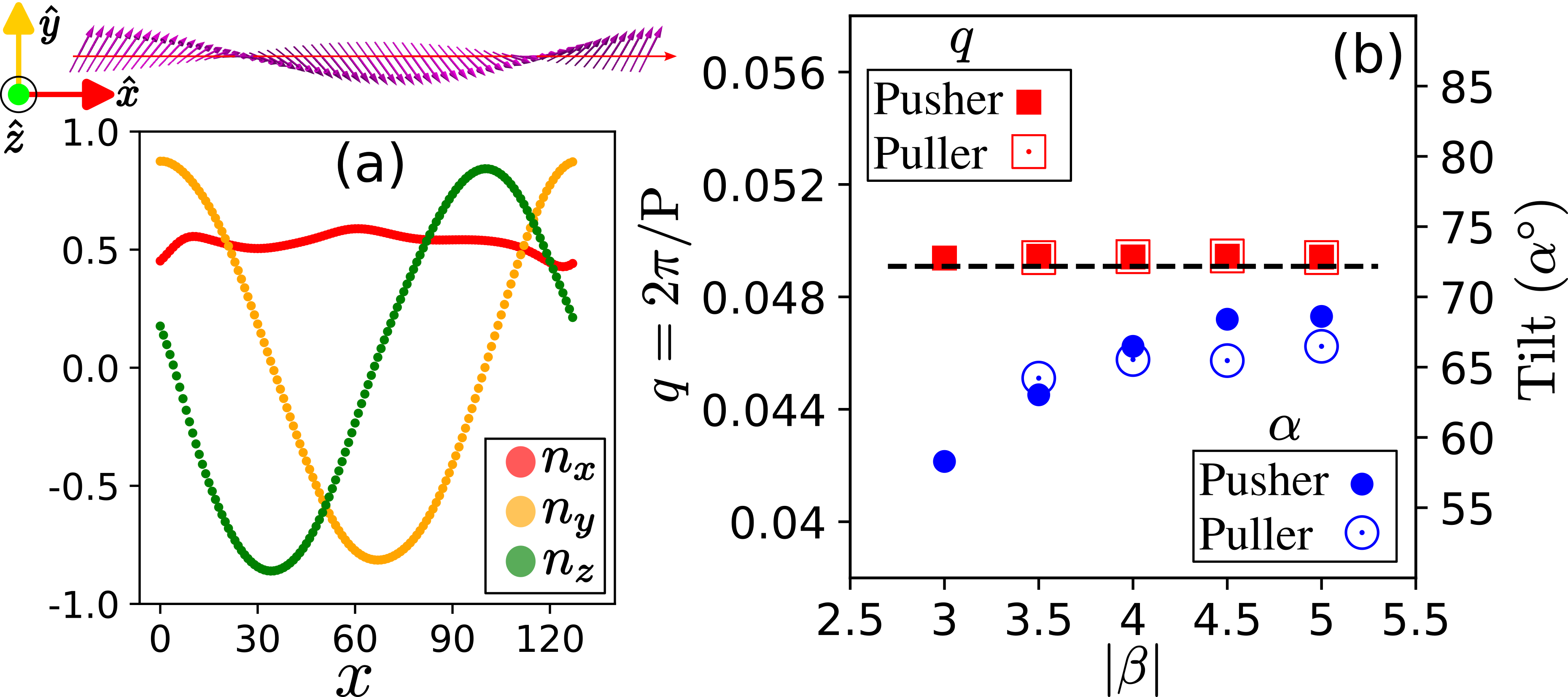}
\caption{(a) An example of the LC director components $n_x$, $n_y$ and $n_z$ along $x$-axis in the helical state. The data can be fitted by director $\hat{\mathbf{n}}$ corresponding to a cholesteric with a $x$ as the helical axis: $n_x=\sin\alpha$, $n_y=\sin\alpha\cos(qx)$ and $n_z=\sin\alpha\sin(qx)$, where $\alpha$  is a tilt angle and $q=2\pi/p$ is an inverse pitch length. (The data corresponds to $\beta = -3.5$ and $\phi = 0.01$). (b) The inverse pitch length $q$ and tilt angle $\alpha$ measured from the simulations as a function of the squirmer parameter $\beta$ at a volume fraction $\phi\approx 1$\%. The horizontal dashed line marks $q\approx 2\pi/L$, where $L$ is the simulation box length.}
\label{fig2}
\end{figure}

\begin{figure} 
\includegraphics[width=1.0\columnwidth]{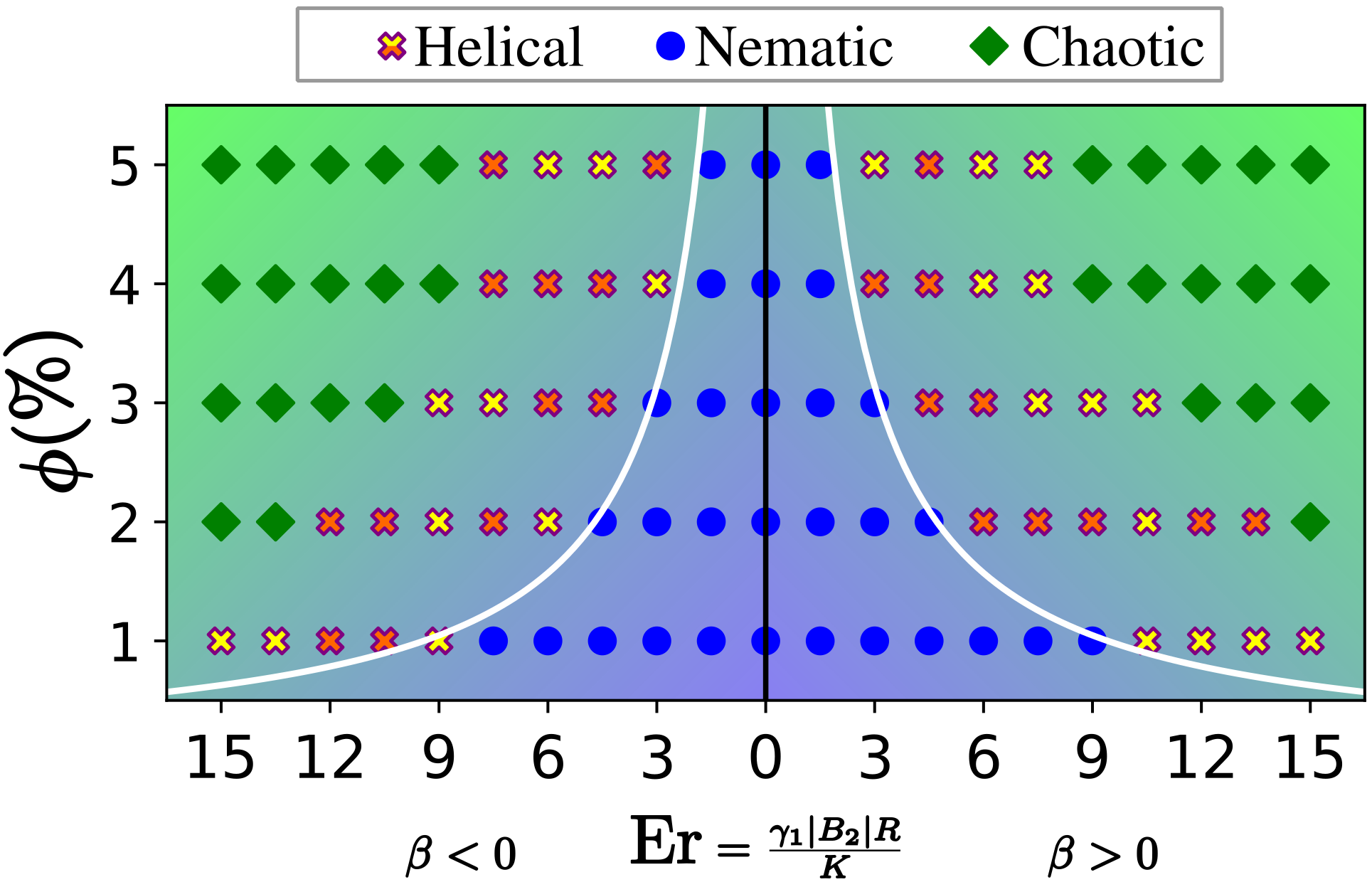}

\caption{Steady state phase diagram for the microswimmer-nematic composite material, as a function {\color{black} of the Ericksen number $\mathrm{Er}=\tfrac{\gamma_1 |B_2| R}{K}$ and the} swimmer volume fraction $\phi$. The blue spheres corresponds to uniform nematic states. The crosses show where the helical states where observed. The purple (yellow) crosses mark the right (left) handed helices. The green diamonds correspond to chaotic states. The critical swimmer volume fraction $\phi^*$ marking the transition between the nematic and helical states is fitted by $\phi^*\sim | B_2 | ^{-1}$ (see text for details).}
\label{fig3}
\end{figure}

Initially, the nematic director $\hat{\mathbf{n}}$ is along $x$-axis (Fig.~\ref{fig1}a). At the onset of the instability, a continuous twist is observed to develop along this axis. The twist has well defined handedness and spans the whole system (Fig.~\ref{fig1}b). However, there is no inherent chirality in the system. Indeed, in the different ensembles, we observed the formation of both left and right handed twists equally (see {\it e.g.} Fig.~\ref{fig3}). 

At the steady state, the $\hat{\mathbf{n}}$ is well fitted with a helical configuration (Fig.~\ref{fig2}a): $n_x=\cos\alpha$, $n_y= \sin\alpha\cos(qx)$ and $n_z=\pm\sin\alpha\sin(qx)$  where $\pm$ corresponds to left and right handed twists, $\alpha$ is the tilt angle respect to $x$-axis and $q$ is an inverse pitch length $q=2\pi/p$. The $q$ is observed to be nearly constant in the helical state for both pushers and pullers, and the pitch length $p$ matches the simulation box length ($p\approx L\approx 21R$; dashed line in Fig.~\ref{fig2}b). The tilt angle $\alpha$ is observed to increase upon increasing the strength of the force dipoles, with the tendency being slightly more pronounced for pushers than pullers (open and closed blue circles in Fig.~\ref{fig2}b).  

The particle trajectories and director orientation are connected, and the particle trajectories become helicoidal in the helical state (Fig.~\ref{fig1}d). The pitch length of the particle trajectories is approximately given by the pitch length of the LC (Fig.~\ref{fig1}b and d). At the steady state, on average, the pushers swim along and pullers perpendicular to the local $\hat{\mathbf n}$, leading to a radius of curvature of the helical trajectory $r_t\approx \tan(\alpha)/q$ and $r_t\approx 1/(q\tan(\alpha))$ for pushers and pullers, respectively. Using the data ($\phi\approx 1$\% and $\beta \approx \pm 3.5$ in Fig.~\ref{fig2}b) these give $r_t\approx 7R$ and $r_t\approx 1.6R$ which agree reasonably with the simulations (dashed and solid lines in the right panel of Fig.~\ref{fig1}d). %corresponding to a slight increase of the overall chirality.

{\color{black} In passive achiral nematics, chiral symmetry breaking have been observed to occur due to externally imposed flow and confinement effects~\cite{vcopar2020microfluidic,vasquez2024control,nayani2015spontaneous,jeong2015chiral,baza2020shear,park2020periodic,sharma2020time,zhang2021structures,zhang2024flow}}.
{\color{black} Here,} the spontaneous formation of the {\color{black} helical states arises from the coupling between the swimmer flow-fields and the nematic director $\hat{\mathbf{n}}$.}
The vorticity $\mathbf{\omega}$ of the squirmer flow-field $\mathbf{v}(\mathbf{r})$ gives rise to a torque on an isolated spherical swimmer in nematic liquid crystals~\cite{lintuvuori2017hydrodynamics}. 
In living liquid crystals thin films~\cite{zhou2014living,genkin2017topological}, a flow instability was shown to arise from the competition between the active (hydrodynamic) torques and elastic aligning torques. We assume similar mechanism here. 

The transition point between the nematic and helical states depends both on the particle volume fraction $\phi$ and the strength of the force dipole {\color{black} $|B_2|$ (Fig.~\ref{fig3}).} 
To phenomenologically relate these quantities to an activity $\zeta$ at the continuum limit, we consider the vorticity $\mathbf{\omega}$ of the squirmer flow-field in isotropic fluid $\mathbf{v}(\mathbf{r})$~\cite{pak2014generalized} at a distance $r$ from another swimmer $\mathbf{\omega}=\nabla\times\mathbf{v}(\mathbf{r})=-3/2\sin 2\theta B_2/r^3\hat{\mathbf{e}_\xi}$, where $\hat{\mathbf{e}_\xi}$ is a unit vector along azimuthal direction. When the density of the particles is uniform, at low $\phi$ the average distance $l$ between the particles follows $l\sim \phi^{-1/3}$. Using these we can  approximate $\zeta~\sim B_2\phi$. When all the other material parameters are unchanged, the instability occurs at a (constant) critical value $\zeta^*$. This gives $\phi^*\sim B_2^{-1}$ for the critical volume fraction, which is in agreement with the predictions for confined 2D living liquid crystals~\cite{genkin2017topological}, and fits the simulation data remarkably well (white lines in Fig.~\ref{fig3}). {\color{black} The onset of the helical state, is observed to happen at moderate Ericksen numbers and span to low swimmer concentrations  $\mathrm{Er}\sim 10$ and $\phi\sim 1$\%, corresponding to  experimentally relevant values~\cite{zhou2014living,zhou2017dynamic}.}

The system is achiral, and we observe an equal amount of left and right handed states (given by yellow and purple crosses in Fig.~\ref{fig3}). This suggests that the chiral symmetry breaking arises from an hydrodynamic instability. In 2-dimensional extensile active nematics~\cite{simha2002hydrodynamic} and in thin-film living liquid crystals~\cite{zhou2014living}, a bend-instability has been observed to be dominant. Our results suggest, that the dominant instability is replaced by a twist when the 3rd dimension is opened. Indeed, {\color{black} linear stability analysis have predicted, a twist-bend mode to be most unstable in  3-dimensional extensile active gels~\cite{chatterjee2021inertia,nejad2022active},} 
{\color{black} and a spontaneous mirror symmetry breaking in the defect dynamics of active nematic gels have been observed both in simulations~\cite{shendruk2018twist} and in experiments~\cite{head2024spontaneous}.}
Very recently a spontaneous flow transition with well  a defined chirality has been predicted in homeotropically confined active nematics~\cite{pratley2023threedimensional}. 

\begin{figure}
\includegraphics[width=0.95\columnwidth]{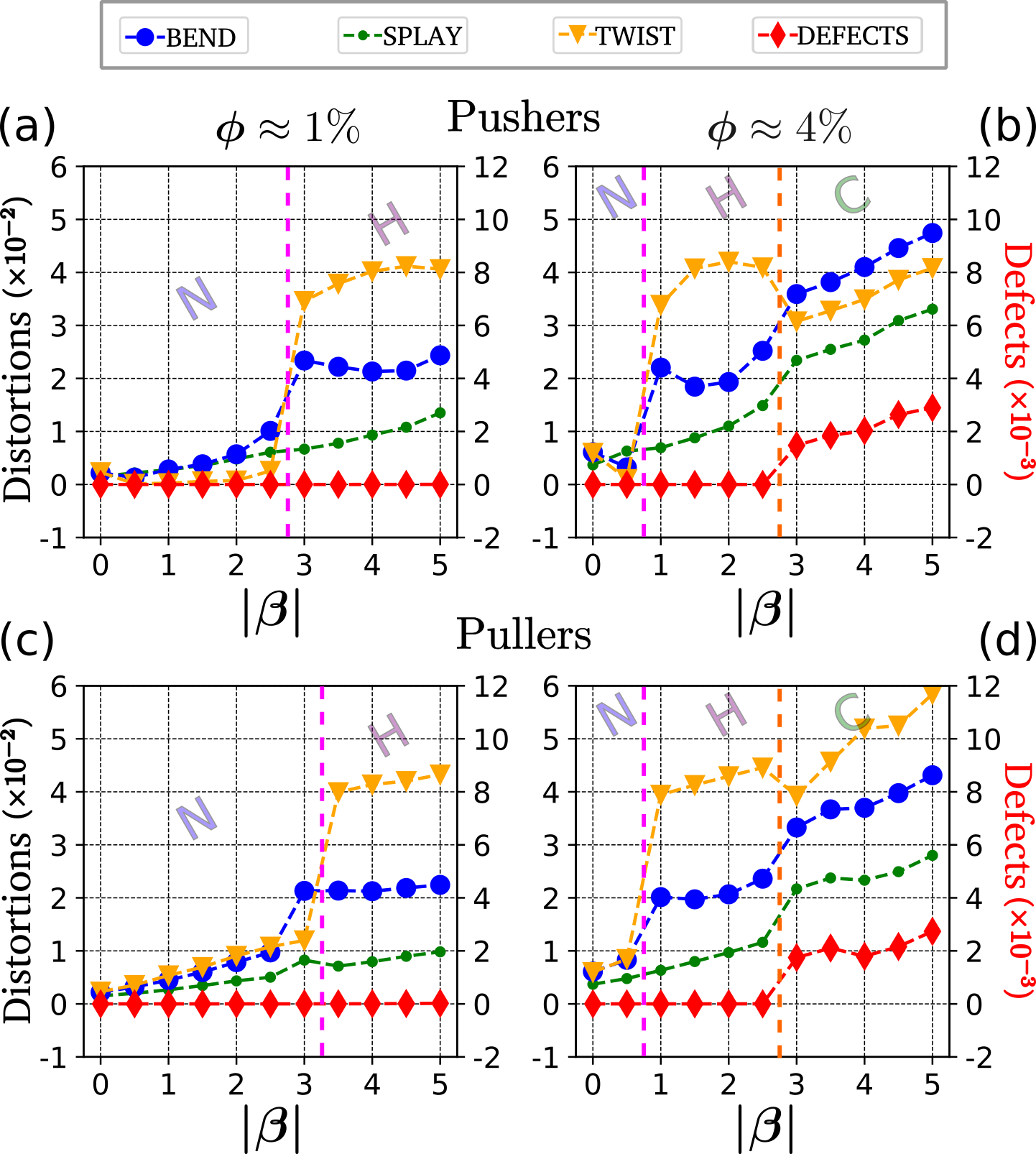}
\caption{The bend $B$ (blue circles), splay $S$ (green dots)  and twist $T$ (yellow triangles) distortions as well as the defect concentration (red diamonds) measured from the simulations for $\phi\approx 0.01$ and $\phi\approx 0.04$ left and right panel, respectively, as a function of the squirmer parameter $\beta$. The dashed vertical pink (orange) lines mark the transition between nematic and helical (helical and chaotic) states.}
\label{fig4}
\end{figure}

In our system, the equilibrium state of the liquid crystal is a uniform nematic. The swimmer flow-fields can perturb this and create (local) deformations, which are penalised by the elastic cost of these distortions. To analyze the different deformations in the system, we calculated the amount of twist, bend and splay~\cite{selinger2018interpretation,SupMat} across nematic, helical and chaotic states (Fig.~\ref{fig4}). In the nematic state, the system has uniform order and any deformations are small and localised near the particles (small $|\beta |$ values in Fig.~\ref{fig4}).  At the onset of the instability, we observe a sudden increase of the distortions. The twist distortions are approximately twice larger than the bend, and four times that of splay, for both pushers and pullers (top and bottom panels in Fig.~\ref{fig4}, respectively). The chaotic state is marked by the emergence of topological defects (Fig.~\ref{fig4}b and d)~\cite{SupMat}. The splay distortion are also observed to grow, while the bend and twist deformations rest largest (large $|\beta|$ values in the Fig.~\ref{fig4}b and d).

The sudden growth of twist and bend distortions at the transition between the uniform nematic and helical states (Fig.~\ref{fig4}a and c), suggests the dominance of the twist-bend mode, in agreement with the linear stability analysis of 3-dimensional extensile active nematics~\cite{chatterjee2021inertia}. The absence of splay instability for pullers, which has been predicted for contractile active nematics~\cite{simha2002hydrodynamic}, can be understood in the terms of the swimming direction of the particles. At the steady state, the pullers swim perpendicular to the (local) nematic director~\cite{lintuvuori2017hydrodynamics}. The perpendicular alignment of an inward (contractile) force-dipole  respect to the LC director, corresponds approximately to a parallel alignment of an outward (extensile) force dipole aligned along $\hat{\mathbf{n}}$. Thus the flow-instability for both spherical pusher and puller swimmers can be expected to be the same, which agrees with our observations (upper and lower panels in Fig.~\ref{fig4}, respectively).

\paragraph*{Conclusions:} Using hydrodynamic simulations, we have studied the collective dynamics of microswimmers in nematic liquid crystal. We observe a spontaneous chiral symmetry breaking, where the uniform nematic order becomes unstable and formation of a continuous twist along a unique axis is observed. The particle dynamics follows the LC order and in the cholesteric state the swimmer trajectories become helical. 

There is no inherent chirality in the system. At the steady state, an equal amount of of left and right handed helices are observed. The chiral states arise from a hydrodynamic instability, originating from the coupling between the swimmer flow-fields and the liquid crystalline elasticity.  By evaluating the distortions in the system, we demonstrate that the dominant mode is a twist-bend instability. This agrees with predictions from a linear stability analysis of 3-dimensional extensile active nematics~\cite{chatterjee2021inertia}. Our predictions could be tested experimentally by opening the 3rd dimension in the experiments of quasi-2d living LCs~\cite{zhou2014living}, {\color{black} where the Ericksen number $\mathrm{Er}\sim 10$ and $\phi\sim 0.2$\%, are commensurate with the parameters considered in our simulations. In these experiments, the lateral size of the system,  is a lot larger than the predicted periodicity $p\sim 21R\sim 140\mu$m, which should allow the helical state to occur.} The predictions for pullers, could be realised by considering, for example, {\it Chlamydomonas} which is near spherical microswimmer with a far-field flow corresponding to a puller force-dipole~\cite{drescher2010direct}.

\begin{acknowledgments}
BG and JSL acknowledge the  French  National  Research  Agency  through  Contract No. ANR-19-CE06-0012  and la region Nouvelle Aquitaine project GASPP for funding, and cluster Curta for computational time.
\end{acknowledgments}

%\end{widetext}
\bibliography{references}

\clearpage

\begin{appendix}

\clearpage

\onecolumngrid

{\center{\large {\bf  Supplementary material for Microswimmers knead nematics into cholesterics} }} \\ 
\medskip
\begin{center}
    Bhavesh Gautam and Juho Lintuvuori
\end{center}

\begin{center}
 Univ. Bordeaux, CNRS, LOMA, UMR 5798, F-33400 Talence, France 
\end{center}

\setcounter{equation}{0}
\setcounter{figure}{0}
\renewcommand{\thefigure}{S\arabic{figure}}
\renewcommand{\theequation}{S\arabic{equation}}
\medskip

\section{Additional details for the calculation of the twist, bend and splay deformations as well as the defect density}
In the simulations, we  use a $\mathbf{Q}$-tensor for the nematic liquid crystal. At each time step, we calculate the Landau-de Gennes free energy density given by equation (1) in the main text, where a single elastic constant approximation is used. % $\frac{K}{2} \left(\partial_{\beta} Q_{\alpha \beta}\right)^2$. %, posing a challenge for evaluating the individual contributions from each distortion.

To evaluate the bend, splay and twist distortions in the system, we consider the  Oseen-Frank formalism and follow the interpretation given in~\cite{selinger2018interpretation}. In this framework, the elastic free energy density $F_{OF}$ associated with the distortions in the uniaxial nematic liquid crystal, with director field $\hat{n}(r)$, is expressed as follows \cite{Lavrentovich_book},
\begin{equation}
    F_{OF} = \frac{1}{2} K_{11} (\nabla \cdot \hat{n})^2 + \frac{1}{2} K_{22} (\hat{n} \cdot \nabla \times \hat{n})^2 + \frac{1}{2} K_{33} (\hat{n} \times (\nabla \times \hat{n}))^2 -K_{24} \nabla \cdot [\hat{n}(\nabla \cdot \hat{n}) + \hat{n} \times (\nabla \times \hat{n})]
    \label{Oseen_franck_full}
\end{equation}
Here, $K_{11}$, $K_{22}$, and $K_{33}$ represents the elastic constants for splay, twist and bend distortions, respectively. The $K_{24}$ is the elastic constant for the fourth type of distortion known as "biaxial splay" \cite{selinger2018interpretation}. %which is not addressed in this paper. 
The splay $S$, bend $\vec{B}$, and twist $T$ distortions are defined as :
\begin{align}
    S=\nabla \cdot \hat{n} \qquad T=\hat{n} \cdot (\nabla \times \hat{n}) \qquad \vec{B}=\hat{n} \times (\nabla \times \hat{n})
\end{align}
While one can use above expressions to calculate the amount of different distortions, it would be advantageous in our simulations to calculate these distortions directly from the $\mathbf{Q}$-tensor. To achieve this, we use the formalism proposed by Selinger~\cite{selinger2018interpretation}, where the distortions are calculated using a tensor $\mathbf{q}$, which is related to $\mathbf{Q}$ by
\begin{equation}
     q_{ij}= \frac{1}{3} \left(\delta_{ij} + \frac{2Q_{ij}}{s}\right).
\end{equation}
$s$ is the scalar order parameter of nematic liquid crystal. Due to inherent ambiguity in defining the splay scalar uniquely in terms of $q_{ij}$, a splay vector $\vec{S}=S\, \hat{n}$ is introdcued and the distortions are given by~\cite{selinger2018interpretation}
\begin{align}
        S_{i}=q_{il} \partial_{j} q_{jl} \qquad     T=\epsilon_{ijk} q_{il} \partial_{j} q_{kl} \qquad     B_{k} = -q_{il} \partial_{i} q_{kl}
\end{align}

We use these expressions to calculate $|\vec{S}|$, $|T|$, and $|\vec{B}|$, representing the magnitudes of splay, bend and twist, respectively,  in the nematic liquid crystal. 
From the \textit{Ludwig} code~\cite{ludwigcode}, the output is a $\mathbf{Q}$-tensor field. These tensors can be diagonalized to extract corresponding eigenvalues and eigenvectors. At each lattice point, the nematic director is determined by taking eigenvector corresponding to the highest absolute eigenvalue. Subsequently, we reconstruct the nematic order tensor using the formula:

\begin{equation}
Q_{ij}= s \left(\frac{3}{2}n_i n_j - \frac{1}{2} \delta_{ij} \right)
\end{equation}
From this, we construct the $\mathbf{q}$ tensor using equation (3), and  calculate distortions using equation (4). The results shown in the Fig. 4 of the main text are averaged over all the liquid crystal lattice points.

To characterise the chaotic LC state, the defect density in the system was calculated. The defects are identified as the regions where the local order parameter $s_{\mathrm{loc}}\leq 0.85 s$, where $s=1/3$ corresponds to perfect nematic order. The figure 4 in the main text, shows the resulting defect  $V_{\mathrm{defect}}/V_{\mathrm{\mathrm{LC}}}$ densities.

\newpage

\section{Additional figures showing particle trajectories and the topological defects in the chaotic state}

\begin{figure}[h]
\includegraphics[width=0.5\columnwidth]{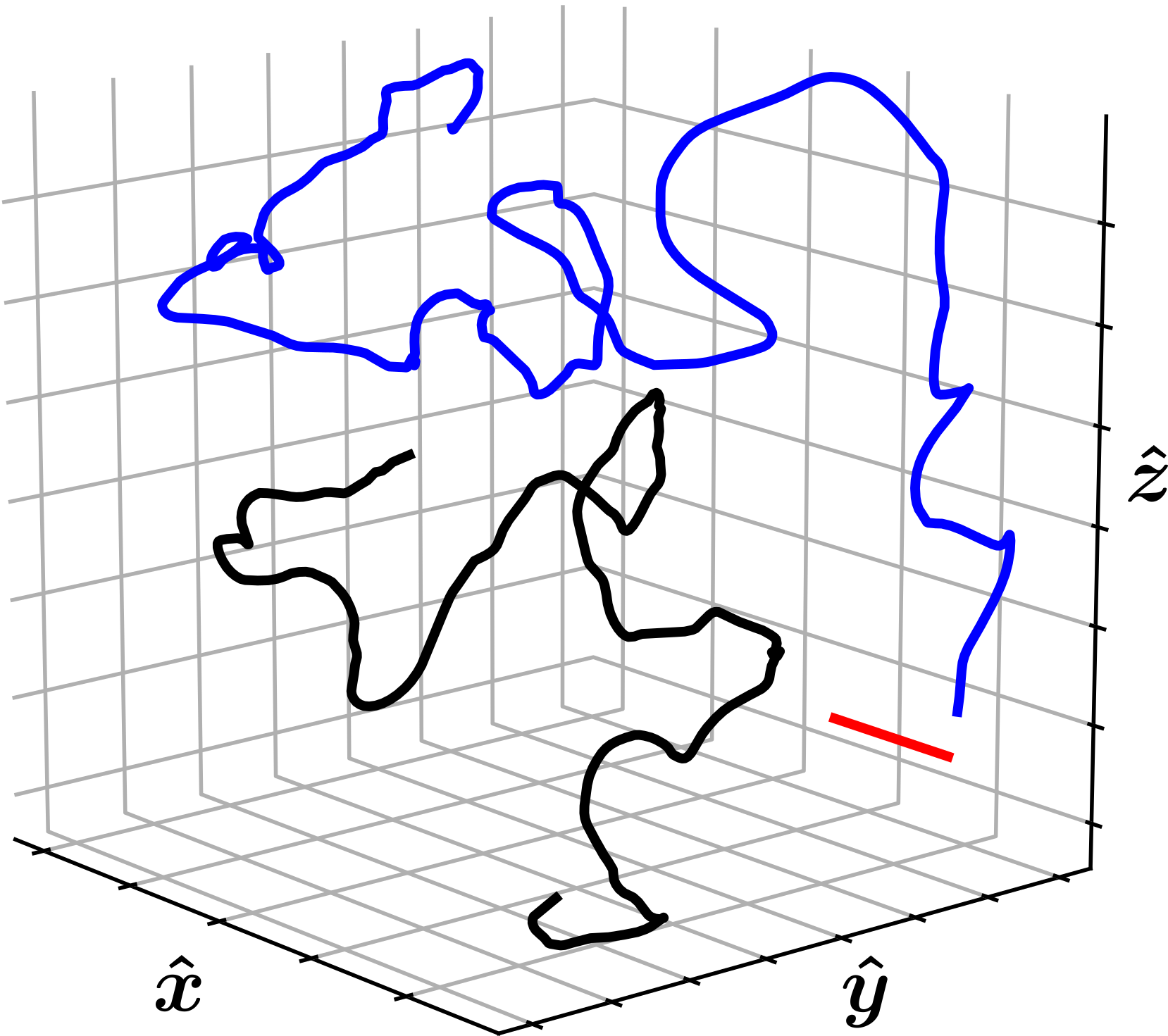}

\caption{Examples of unwrapped trajectories of pushers in the chaotic state are given by blue and black solid lines. In the chaotic state, the pushers align themselves, on average, with the local nematic director, but its trajectory lacks a specific pattern due to spatial variations in the director field. The red scale bar corresponds to the system size $L\approx 21R$ and the data to $\phi \approx 4\%$ and $\beta \approx -3.5$ ($\mathrm{Er}\approx 10.5$)}
\label{fig_chaotic_traj}
\end{figure}

\begin{figure}[h] 
\includegraphics[width=0.5\columnwidth]{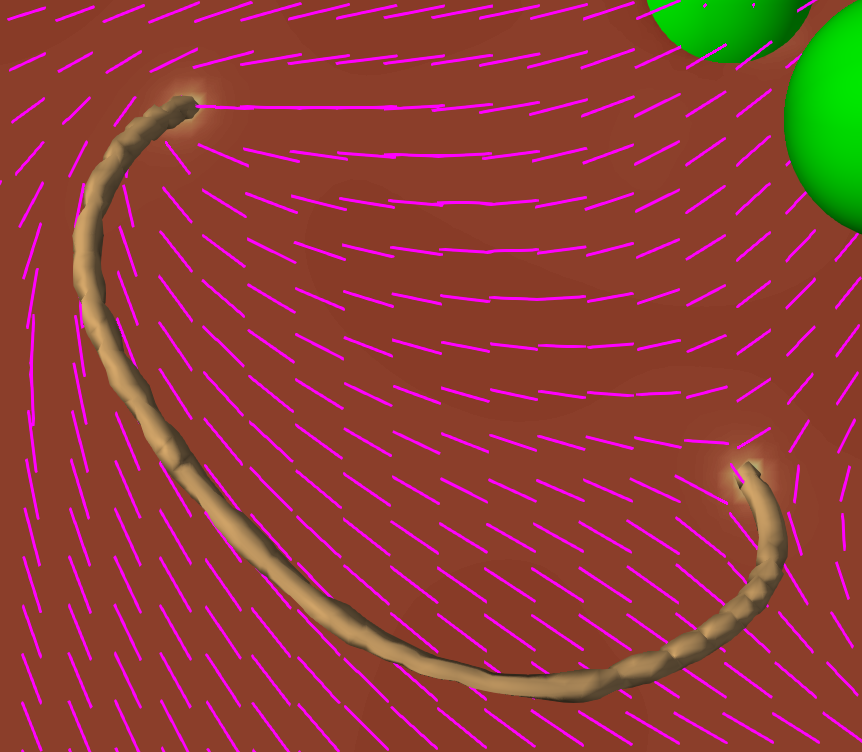}
\caption{An example of the topological defects observed in the chaotic state. $+1/2$ defect top left, and $-1/2$ bottom right. The brown ribbon corresponds to the disclination line in 3-dimensions. The simulations corresponds to $\phi\approx 2$\% and $\beta = -5$ ($\mathrm{Er}\approx 15$.)} 
\end{figure}

%\newpage

%\bibliography{references}
\end{appendix}
\end{document}